\documentclass[pre,aps,twocolumn,superscriptaddress,showpacs]{revtex4}

\usepackage{epsfig,amsmath,amssymb}
\newcommand{\rv}{{\bf r}}
\newcommand{\xv}{{\bf x}}
\newcommand{\yv}{{\bf y}}

\newcommand{\J}{{\sf J}}
\newcommand{\M}{{\sf M}}
\newcommand{\N}{{\sf N}}
\newcommand{\K}{{\sf K}}
\newcommand{\G}{{\sf G}}
\newcommand{\phizero}{\phi_{\rm 0d}}

\begin{document}

\title{Density functional for ternary non-additive hard sphere mixtures}

\author{Matthias Schmidt}
\affiliation{Theoretische Physik II, Physikalisches Institut, 
  Universit{\"a}t Bayreuth, D-95440 Bayreuth, Germany}
\affiliation{H. H. Wills Physics Laboratory, University of Bristol, 
  Royal Fort, Tyndall Avenue, Bristol BS8 1TL, United Kingdom}

\date{29 July 2011, revised version: 31 August 2011, 
  J. Phys.: Condensed Matter {\bf 23}, 415101 (2011).}

\begin{abstract}
  Based on fundamental measure theory, a Helmholtz free energy density
  functional for three-component mixtures of hard spheres with
  general, non-additive interaction distances is constructed. The
  functional constitutes a generalization of the previously given
  theory for binary non-additive mixtures. The diagrammatic structure
  of the spatial integrals in both functionals is of star-like (or
  tree-like) topology. The ternary diagrams possess a higher degree of
  complexity than the binary diagrams.  Results for partial pair
  correlation functions, obtained via the Ornstein-Zernike route from
  the second functional derivatives of the excess free energy
  functional, agree well with Monte Carlo simulation data.
\end{abstract}


\maketitle
\section{Introduction}
Rosenfeld's fundamental measures theory (FMT) for additive hard sphere
mixtures \cite{rosenfeld89,kierlik90} has become a cornerstone of
classical density functional theory (DFT) \cite{evans79}. FMT features
in numerous applications to a wide variety of interesting phenomena in
liquids \cite{roth10review,tarazona08review,lutsko10review}. In
additive hard sphere mixtures the interaction distance between unlike
components is taken to be the arithmetic mean of the (like-species)
diameters.  Additive mixtures are often considered as prototypical in
the description of liquid mixtures. Nevertheless, non-additivity is a
generic feature \cite{kahl90} that arises very naturally in effective
interactions, e.g.\ due to the depletion effect \cite{louis01depl}, or
when integrating out solvent degrees of freedom in electrolytes
\cite{kalcher10,kalcher10jcp}. A generalization of FMT to binary
non-additive hard sphere (NAHS) mixtures \cite{schmidt04nahs} was
based on the scalar version of the additive hard sphere functional
\cite{kierlik90}, and has been used successfully in the investigation
of bulk \cite{hopkins10nahs,ayadim10,hopkins11natpl} and of
interfacial \cite{hopkins11nawe} phenomena that occur in NAHS
mixtures.

Recently significant progress has been made in the formalization of
the mathematical structure of FMT. This includes i) insights into the
geometry of the non-local aspects of the theory, i.e.\ the algebraic
group structure and symmetry properties of the convolution kernel
matrix $\K(R,r)$ \cite{schmidt04nahs,schmidt07nage,schmidt11nagl},
which controls the range of non-locality in the functional. Here $R$
is a fixed lengthscale and $r$ is the radial distance in
three-dimensional space. Formally, $\K(R,r)$ is a $4\times 4$-matrix
that is indexed by powers of lengthscale. Its remarkable algebraic
properties \cite{schmidt07nage} allow to view it as an object that is
suitable to add or remove a layer of thickness~$R$ from a given
sphere.  Furthermore, ii) the FMT for additive hard sphere mixtures
was obtained from a tensorial-diagrammatic series in density
\cite{leithall11hys}.  The diagrams possess star-like topology, which
is an approximation of the combinatorial complexity of the exact
virial expansion. Apart from a central space integral, all field
points are integrations over the density field(s), as they are in the
exact virial expansion. The bonds, however, are weight functions
rather than Mayer functions, and possess only half the range of the
(hard sphere) Mayer function.  In one dimension, the result from the
series is equal to Percus' exact functional \cite{percus76}. In three
dimensions it gives the Kierlik-Rosinberg form \cite{kierlik90} of
FMT. The five-dimensional hard hypersphere version is investigated in
detail in Ref.\ \cite{leithall11hys}. Comparison to data from the
literature for bulk structure and thermodynamics demonstrates the
capability of this theory for study bulk and inhomogeneous hypersphere
mixtures.

In the present work we generalize the tensorial-diagrammatic series to
non-additive hard sphere interactions. We use the kernel matrix
$\K(R,r)$ as a further type of bond. This enables us to represent the
binary NAHS functional of Ref.\ \cite{schmidt04nahs} as a series of
diagrams that are formed by two stars, one for each species. Here the
center of one star is connected to the center of the second star by a
$\K$-bond. Formulating an FMT for general ternary non-additive hard
sphere mixtures requires to modify this topology.  Treating special
cases of ternary mixtures that have a suitably high degree of
additivity amounts to straightforward generalizations of
Ref.\ \cite{schmidt04nahs}. This is the case when the cross
interaction between (say) species 1 and 2 is additive, and the
non-additivities between 13 and 23 are coupled in a certain way (see
the discussion below (\ref{EQone})), and hence cannot be chosen
independently from each other.  The general case, however requires the
introduction of diagrams with different topology.  Here three stars,
one for each species, are connected to a central three-arm star.  As
laid out in detail below, all (four) inner junctions are bare space
integrals, without multiplication by a one-body density.  Only the
outer ends carry multiplication by a (bare) density variable.  We show
that the theory predicts the bulk structure of ternary mixtures, via
the Ornstein-Zernike route, with good quality as compared to Monte
Carlo simulation data.

\section{Non-additive hard sphere interactions}
Non-additive hard sphere mixtures possess pair potentials $v_{ij}(r)$
between species~$i$ and $j$ as a function of the center-center
distance $r$ of the two particles, that are given as
$v_{ij}(r)=\infty$ if $r<\sigma_{ij}$, and zero otherwise. In the
general case all $\sigma_{ij}$ are independent of each other, except
for the trivial symmetry $\sigma_{ij}=\sigma_{ji}$.  Non-additivity
parameters are conventionally defined as
$\Delta_{ij}=2\sigma_{ij}/(\sigma_{ii}+\sigma_{jj})-1$ for $i\neq j$,
where $\Delta_{ij} \geq -1$.  For additive mixtures all
$\Delta_{ij}=0$. For positive non-additivity, $\Delta_{ij}>0$, the
unlike components interact at a larger distance than the arithmetic
mean of their diameters. For negative non-additivity, $\Delta_{ij}<0$,
the interaction distance is smaller than the mean of the diameters.
While a binary NAHS mixture is characterized by either positive or
negative non-additivity, in ternary systems mixed cases are possible,
where not all of the $\Delta_{ij}$ have the same sign.  In ternary
NAHS mixtures the equation of state \cite{santos05} was considered and
phase stability was investigated using integral equation theory
\cite{gazzillo96}. Phase equilibrium was also considered in
polydisperse non-additive hard sphere systems
\cite{dickinson79,paricaud08}.

\section{Density functional theory}
\subsection{Overview and choice of lengthscales}
\begin{figure}[htbp]
\centering
\includegraphics[width=8cm]{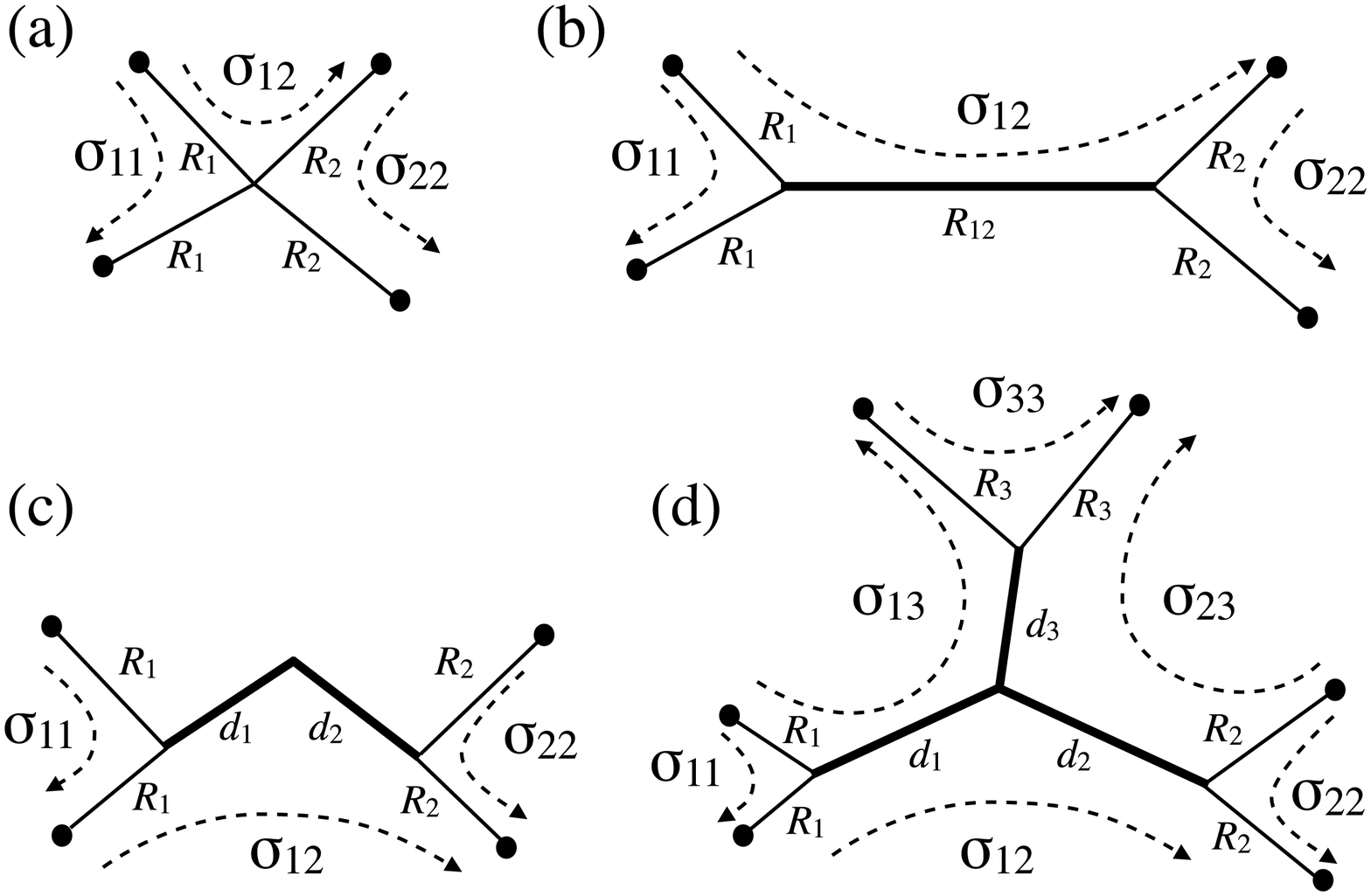}
\caption{Illustration of the decomposition (\ref{EQone}) of the hard
  sphere interaction distances $\sigma_{ij}$ into like-species
  particle radii $R_i$ and non-additivity distances $d_i$. Thin solid
  lines represent distances between a space point (dots) and a
  junction; thick solid lines represent the distance between two
  junctions.  The dashed arrows indicate paths that possess
  length~$\sigma_{ij}$ (as indicated); the sense of direction is only
  a guide to the eye. Decompositions are shown for the binary additive
  mixture~(a), the binary non-additive mixture with the cross diameter
  being divided into three~(b) and four~(c) contributions, and the
  ternary non-additive mixture~(d). The two binary cases (b) and (c)
  are equivalent when $R_{12}=d_1+d_2$.  
}
\label{FIGone}
\end{figure}

In order to construct an FMT for ternary NAHS mixtures, we first
identify suitable lengthscales. Consistent with Rosenfeld's additive
case \cite{rosenfeld89}, we use the particle radii $R_i=\sigma_{ii}/2$
of each species $i=1,2,3$. For the binary mixture
\cite{schmidt04nahs}, the cross diameter between species 1 and 2 was
decomposed as $\sigma_{12}=R_1+R_{12}+R_2$, where the lengthscale
$R_{12}=\sigma_{12}-(\sigma_{11}+\sigma_{22})/2$ accounts for the
non-additivity.  For the ternary system, we generalize this to a
decomposition of the cross diameters into sums of {\em four}
contributions,
\begin{align}
  \sigma_{ij}&=R_i+d_i+d_j+R_j,
  \label{EQone}
\end{align}
where $ij=12,13,23$, and the three lengthscales $d_i$, $i=1,2,3$,
control the degrees of non-additivity. Inverting Eq.~(\ref{EQone}) is
possible for ternary mixtures and yields
$d_j=(\sigma_{ij}+\sigma_{jk}-\sigma_{ik}-\sigma_{jj})/2$; here
$ijk=123$ or any permutation thereof. As a simple check, a counting
exercise assures us that the number of parameters $R_i$ (three) and
$d_i$ (also three) is enough to represent the six independent
components of $\sigma_{ij}$.  The binary case above is recovered if we
set $R_{12}=d_1+d_2$. Here the relative splitting of $R_{12}$ into
$d_1$ and $d_2$ is arbitrary.  If we keep three species, and set
$d_1=d_2=0$, then only $d_3\neq 0$ remains in order to control the
non-additivities between 13 and between 23. For mixtures with such
restricted degree of non-additivity, as mentioned above, the binary
NAHS functional can be generalized easily.  Figure~\ref{FIGone}
illustrates the different types of decomposition of the $\sigma_{ij}$
in binary additive (a), binary non-additive (b, c), and ternary
non-additive (d) mixtures. Note that the $d_i$ can be negative, but
that $R_i+d_i+d_j+R_j\geq 0$ must hold due to
(\ref{EQone}). Furthermore certainly $R_i\geq 0$.

As a means to control the different lengthscales in the density
functional, we use the kernel matrix $\K(R,r)$ of
Refs.\ \cite{schmidt04nahs,schmidt07nage,schmidt11nagl}.  Two of its
properties render this a suitable object for the construction of the
density functional: i) the four scalar Kierlik-Rosinberg weight
functions (where $R$ is identified with the particle radius) feature
as components and ii) matrices can be chained,
$\K(R+R',r)=\K(R,r)\ast\K(R',r)$, where the asterisk denotes the
three-dimensional spatial convolution and matrix multiplication is
implied on the right hand side. The three-dimensional Fourier
transform of $\K(R,r)$ can be expressed as $\tilde \K(R,q)=\exp(R\G)$,
where $q$ is the radial distance in reciprocal space, and the
(generator) matrix $\G$ depends on $q$ and is defined by its
components $G_1^{\;\;0}=1, G_2^{\;\;1}=8\pi, G_3^{\;\;2}= 1,
G_1^{\;\;2}=-q^2/(4\pi), G_0^{\;\;3}=-q^4/(8\pi)$; all other
$G_\mu^{\;\;\nu}=0$. Here the lower index indicates the row and the
upper index indicates the column; all Greek indices run from 0 to 3
here and in the following. Due to the symmetry
$K_\mu^{\;\;\nu}(R,r)=K_{3-\mu}^{\;\;3-\nu}(R,r)$
\cite{schmidt11nagl}, the matrix $\K(R,r)$ has ten independent
components \cite{hopkins10nahs,schmidt11nagl}. The four scalar
Kierlik-Rosinberg \cite{kierlik90} weight functions $w_\nu(R,r)$ are
contained herein, i.e.\ $K_0^{\;\;\nu}(R,r)=w_\nu(R,r)$, explicitly
given in real space as $w_3(R,r)=\Theta(R-r)$, $w_2(R,r)=\delta(R-r)$,
$w_1(R,r)= \delta'(R-r)/(8\pi)$, and
$w_0(R,r)=-\delta''(R-r)/(8\pi)+\delta'(R-r)/(2\pi r)$. An explicit
real-space expression for all further components of $\K(R,r)$ can be
found in \cite{schmidt07nage}.

The Mayer $f$-bond for hard spheres equals $-1$ if the two spheres
overlap and vanishes otherwise. For like species the deconvolution
into scalar weight functions \cite{kierlik90} can be written as
$f_{ii}(r)=-\sum_{\mu=0}^3 w_\mu(R,r) \ast w_{3-\mu}(R,r)\equiv
-\sum_{\mu,\nu=0}^3 w_\mu(R,r)M^{\mu\nu}w_\nu(R,r)$. Here the spatial
convolution of two functions is defined as $(f\ast g)(\xv)=\int d\rv
f(\rv)g(\xv-\rv)$, and the metric $\M$ \cite{schmidt11nagl} has
components $M_{\mu\nu}=M^{\mu\nu}=1$ if $\mu+\nu=3$ and is zero
otherwise. The crucial property of $\K(R,r)$ that we will exploit in
the following is the group structure for the combined operation of
matrix multiplication and (real-space) convolution. For the
interactions between particles of the same species
$\K(2R_i,r)=\K(R_i,r)\ast\K(R_i,r)$, where the convolution product
implies also matrix multiplication, and the Mayer bond is just the
special case $f_{ii}(r)=-w_3(2R_i,r)=-K_0^{\;\;3}(2R_i,r)$. We exploit
the fact that several matrices can be chained \cite{schmidt07nage}, in
order to model the interactions between unlike species via
$\K(\sigma_{ij},r) = \K(R_i,r) \ast \K(d_i,r) \ast \K(d_j,r) \ast
\K(R_j,r)$, where the lengthscales on the right hand side satisfy
(\ref{EQone}). Furthermore the terms on the right hand side commute
(i.e.\ the group is Abelian \cite{schmidt07nage}).  The Mayer bond
between unlike species $i$ and $j$ is
$f_{ij}(r)=-K_0^{\;\;3}(\sigma_{ij},r)$ and can hence be written as
$f_{ij}(r)=-\sum_{\mu,\mu',\nu,\tau}^3 M^{\mu\mu'}w_{\mu'}(R_i,r)\ast
K_\mu^{\;\;\nu}(d_i,r) \ast K_\nu^{\;\;\tau}(d_j,r) \ast
w_\tau(R_j,r)$. This identity, as well as that for the intra-species
case above, can be verified by explicit algebra, most conveniently in
Fourier space, where the convolutions become mere products and
algebraic theorems for trigonometric functions can be used to simplify
the expressions.

\subsection{Algebraic-diagrammatic structure of the free energy functional}
\begin{figure}[htbp]
\centering
\includegraphics[width=8cm]{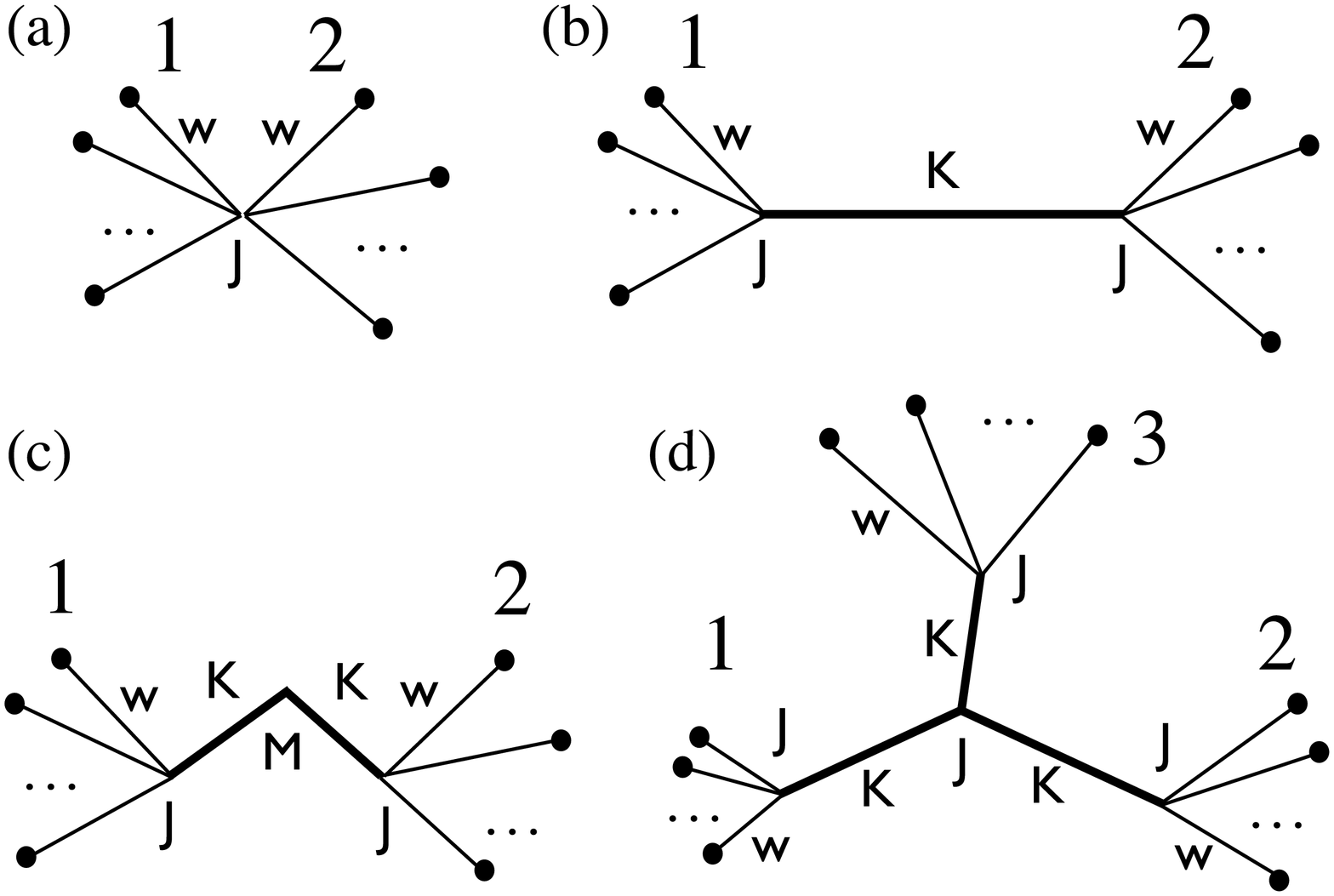}
\caption{Illustration of the topology of the diagrams that constitute
  the density series of the FMT free energy functional for various
  hard sphere mixtures. The one-body density distribution of each
  species $\rho_i(\rv)$ (represented by a filled symbol) is connected
  by weight function bonds, ${\sf w}=(w_0,w_1,w_2,w_3)$ (thin lines),
  to a central space integral (junction).  The different topologies
  are for the binary additive hard spheres (a), binary non-additive
  hard spheres (b,c) and ternary non-additive hard sphere mixtures
  (d). The numbers indicate the different species. In (b,c,d) the the
  junctions are joined by convolution kernel bonds $\K$. The bonds
  that meet at a junction are multiplied by third-rank tensors
  $\J$. Multiplying with the metric $\M$ at the center of (c) joins
  the two $\K$-bonds and restores the topology of (d). All junctions
  and all end points are integrated over.}
\label{FIGdiagrams}
\end{figure}

Using the four weight functions $w_\mu(R_i,r)$, we build
species-dependent weighted densities $n_\mu(i,\rv)$ in the standard
way via convolution with the bare density distribution of the
corresponding species, 
\begin{align}
  n_\mu(i,\rv) = w_\mu(R_i,r) \ast \rho_i(\rv).
\end{align}
We use the third-rank ``junction'' tensor $\J$ of
Ref.~\cite{leithall11hys} in order to couple the $n_\mu(i,\rv)$ and
hence generate terms that are non-linear in densities. Let us denote
the components of $\J$ by $J^{\mu\nu\tau}$. The tensor is symmetric
under exchange of indices, and is non-zero only if
$\mu+\nu+\tau=6$. One can specify $\J$ completely via the elements
$J^{123}=J^{033}=1$, and $J^{222}=1/(4\pi)$.  As a basic building
block for the construction of the density functional, we use the
matrix $\N(i,\rv)$ of weighted densities \cite{leithall11hys} that is
obtained by contracting the vector of weighted densities
$n_\tau(i,\rv)$ with the $\J$-tensor and lowering one of the indices
via contraction with the metric $\M$. Hence the components of the
matrix $\N(i,\rv)$ are obtained as
$N_\mu^{\;\;\nu}(i,\rv)=\sum_{\mu',\tau=0}^3
M_{\mu\mu'}J^{\mu'\nu\tau}n_\tau(i,\rv)$, and given explicitly by
\begin{align}
  \N(i,\rv) &=  \left(\begin{matrix}
    n_3(i,\rv) & n_2(i,\rv) & n_1(i,\rv) & n_0(i,\rv)\\
    0 & n_3(i,\rv) & \frac{n_2(i,\rv)}{4\pi} & n_1(i,\rv)\\
    0 & 0 & n_3(i,\rv) & n_2(i,\rv)\\
    0 & 0 & 0 & n_3(i,\rv)
  \end{matrix}\right).
\end{align}
As an illustration of the power of this formalized framework, we can
obtain \cite{leithall11hys} the additive FMT functional as the
03-component of $\int d\xv \phizero(\sum_i\N(i,\xv))$, where the
integration variable was renamed from $\rv$ to $\xv$, and the
zero-dimensional excess free energy is
\begin{align}
  \phizero(\eta)\equiv(1-\eta)\ln(1-\eta)+\eta =\sum_{m=2}^\infty
  \frac{\eta^m}{m(m-1)},
\end{align}
with the (dummy) variable $\eta$ being the average occupation number
of the zero-dimensional system. Here a function of a matrix is defined
via its power series. In order to represent the mathematical structure
of the integrals in the density functional, we use the diagrammatic
formulation of Ref.\ \cite{leithall11hys}, see Fig.~\ref{FIGdiagrams}
for an overview. In particular, the star topology shown in
Fig.~\ref{FIGdiagrams}a constitutes the relevant type of diagram for
additive hard sphere mixtures. The center of the diagram represents
the integration variable $\xv$, the arms represent the weight
functions $w_\mu(r)$, and the filled symbols represent the one-body
density $\rho_i(\rv)$ at space point(s)~$\rv$. All spatial variables
are integrated over. The number of arms equals the order in
density. Summing up all orders and using the coefficients of the power
series of $\phizero$ yields the Kierlik-Rosinberg free energy
functional~\cite{leithall11hys}.

In the following we require only the last column of the $l$-th power
of the density matrix, obtained formally as
\begin{align}
  \psi^{(l)}_\nu(i,\rv) &= \N^l(i,\rv) \cdot (0,0,0,1)^{\rm t},
  \label{EQpsiPower}
\end{align}
where $\N^l(i,\rv)$ indicates the $l$-th (matrix) power of
$\N(i,\rv)$, the dot represents the multiplication between a matrix
and a vector and the superscript t indicates
transposition. Explicitly, the four components in (\ref{EQpsiPower}) are
\begin{align}
   \psi_0^{(l)}(i,\rv) &= l n_0 n_3^{l-1} - l(l-1)n_1 n_2 n_3^{l-2} \nonumber\\
   &\quad\quad
    +\frac{l(l-1)(l-2)}{24\pi} n_2^3 n_3^{l-3} \label{EQpsi0}\\
   \psi_1^{(l)}(i,\rv) &= l n_1 n_3^{l-1}-\frac{l(l-1)}{8\pi} n_2^2 n_3^{l-2}\\
   \psi_2^{(l)}(i,\rv) &= l n_2 n_3^{l-1}\\
   \psi_3^{(l)}(i,\rv) &= n_3^l, \label{EQpsi3}
\end{align}
where the superscripts of the weighted densities indicate (scalar)
powers, the weighted densities are those of species $i$,
i.e.\ $n_\mu\equiv n_\mu(i,\rv)$, and the arguments have been omitted
for clarity.

For non-additive mixtures we ``transport'' the expressions
(\ref{EQpsi0})-(\ref{EQpsi3}) via convolution with $\K(d_i,r)$. We
hence obtain a fourvector for each species $i$, indexed by
$\mu=0,1,2,3$, and given as
\begin{align}
  \Phi_\mu^{(l)}(i,\rv)&= \sum_{\nu=0}^3\K_\mu^{\;\;\nu}(d_i,r)\ast
  \psi_\nu^{(l)}(i,\rv).
  \label{EQphimu}
\end{align}
These objects serve as ansatz functions for representing the free
energy density; they are specific for each species (indicated by the
argument $i$), are of $l$-th order in density, and carry the dimension
of (length)$^{3-\mu}$.

\subsection{Rewriting the binary non-additive hard sphere functional}
Using the above definitions, we can
write the density functional for {\em binary} non-additive hard sphere
mixtures~\cite{schmidt04nahs} as
\begin{align}
F_{\rm exc}[\rho_1,\rho_2]&=k_BT
\int d\yv \sum_{\mu,\nu=0}^3 M^{\mu\nu}
  \sum_{k,l=0}^\infty \frac{(k+l-2)!}{k!\,l!}
\nonumber\\&\qquad\qquad\times
  \Phi_\mu^{(k)}(1,\yv)\Phi_\nu^{(l)}(2,\yv),
  \label{EQFexcBinarySeries}
\end{align}
where we take the convention that the factorial vanishes for negative
arguments and we have renamed the spatial integration variable from
$\rv$ to $\yv$. The scalar coefficients inside of the double sum over
$k,l$ are those in the Taylor expansion of the zero-dimensional excess
free energy, which for a binary mixture is $\phizero(\eta_1+\eta_2)=
\sum_{k,l=1}^\infty
\binom{k+l}{k}\eta_1^k\eta_2^l/[(k+l)(k+l-1)]$. Writing the
coefficients as $\binom{k+l}{k}/[(k+l)(k+l-1)] =(k+l-2)!/(k!\,l!)$
gives the form in~(\ref{EQFexcBinarySeries}).

A closed expression for the series (\ref{EQFexcBinarySeries}) can be
obtained. This is identical to the previously given
\cite{schmidt04nahs} form of the binary functional
\begin{align}
  F_{\rm exc}[\rho_1,\rho_2] = k_BT\nonumber &\\ \times
  \int d\xv \int d \xv' & \sum_{\mu,\nu=0}^3 K^{\mu\nu}(d_1+d_2,\xv-\xv')
  \Phi_{\mu\nu}(\xv,\xv'),
  \label{EQFexcBinaryExplicit}
\end{align}
where the components of the free energy tensor are
\begin{align}
  \Phi_{\mu\nu}(\xv,\xv')&=
  \sum_{\alpha,\beta=0}^3 
  A_{\mu\alpha}(1,\xv)A_{\nu\beta}(2,\xv')
  \nonumber\\&\qquad\times
  \phizero^{[\alpha+\beta]}\big(n_3(1,\xv)+n_3(2,\xv')\big),
  \label{EQ_Phi_munu}
\end{align}
where $\phizero^{[\alpha]}(\eta)\equiv d^\alpha \phizero(\eta)/d
\eta^\alpha$ is the $\alpha$-th derivative of the $0d$ excess free
energy. Explicit expressions for the coefficients in
(\ref{EQ_Phi_munu}) are
\begin{align}
  A_{01}(i,\xv) &= n_0(i,\xv),\nonumber\\
  A_{02}(i,\xv) &= n_1(i,\xv) n_2(i,\xv),\nonumber\\
  A_{03}(i,\xv) &= \frac{[n_2(i,\xv)]^3}{24\pi},\nonumber\\
  A_{11}(i,\xv) &= n_1(i,\xv),\label{EQAterms}\\
  A_{12}(i,\xv) &= \frac{[n_2(i,\xv)]^2}{8\pi},\nonumber\\
  A_{21}(i,\xv) &= n_2(i,\xv),\nonumber\\
  A_{30}(i,\xv) &= 1,\nonumber
\end{align}
where $i$ labels the species [$i=1,2$ for (\ref{EQ_Phi_munu})]. In
the above we have exploited the convolution property
\begin{align}
  K^{\mu\nu}(d_1+d_2,\xv-\xv')= \nonumber & \\
  \sum_{\tau,\tau'=0}^3  M_{\tau\tau'} \int
  d\yv & K^{\mu\tau}(d_1,\xv-\yv) K^{\tau'\nu}(d_2,\xv'-\yv).
\end{align}

Due to the group structure of the convolution kernels
\cite{schmidt07nage,schmidt11nagl}, the identity
$K_\mu^{\;\;\nu}(d_i+d_j,r)= \sum_{\tau=0}^3
K_\mu^{\;\;\tau}(d_i,r)\ast K_\tau^{\;\;\nu}(d_j,r)$ holds, such that
the lengthscale $R_{12}\equiv\sigma_{12}-(\sigma_{11}+\sigma_{22})/2$
of the binary functional \cite{schmidt04nahs} is recovered as
$R_{12}=d_1+d_2$.  The structure of the diagrams corresponding to
(\ref{EQFexcBinaryExplicit}) is shown in Fig.\ \ref{FIGdiagrams}b and
that corresponding to (\ref{EQFexcBinarySeries}) is shown in
Fig.\ \ref{FIGdiagrams}c.

\subsection{Constructing a ternary non-additive hard sphere functional}
The benefit of re-writing the binary functional in the form
(\ref{EQFexcBinarySeries}) is that this allows for straightforward
generalization to three-component mixtures as
\begin{align}
  & F_{\rm exc}[\rho_1,\rho_2,\rho_3]=  k_BT  \int d\xv 
  \sum_{\mu,\nu,\tau=0}^3   J^{\mu\nu\tau} \nonumber\\ & \times
  \sum_{k,l,m=0}^\infty \frac{(k+l+m-2)!}{k!\,l!\,m!}
  \Phi_\mu^{(k)}(1,\xv)\Phi_\nu^{(l)}(2,\xv)\Phi_\tau^{(m)}(3,\xv),
  \label{EQFexcTernarySeries}
\end{align}
where $(k+l+m-2)!/(k!\,l!\,m!)$ is the coefficient of order
$\eta_1^k\eta_2^l\eta_3^m$ in the Taylor expansion of
$\phizero(\eta_1+\eta_2+\eta_3)$. Again a closed expression can be
obtained, which is in the form
\begin{widetext}
\begin{align}
  F_{\rm exc}[\rho_1,\rho_2,\rho_3] = & k_BT
   \int d\yv \int d\xv \int d\xv' \int d\xv''
   \sum_{\mu,\nu,\tau=0}^3  J^{\mu\nu\tau}
   \sum_{\mu',\nu',\tau'=0}^3 
   K_\mu^{\;\;\mu'}(d_1,\xv-\yv)
   \nonumber \\ & \quad\quad\times
   K_\nu^{\;\;\nu'}(d_2,\xv'-\yv)
   K_\tau^{\;\;\tau'}(d_3,\xv''-\yv)
   \Phi_{\mu'\nu'\tau'}(\xv,\xv',\xv''),
   \label{EQFexcTernaryExplicit}
\end{align}
\end{widetext}
where
\begin{align}
  &\Phi_{\mu\nu\tau}(\xv,\xv',\xv'') = \nonumber\\ & \sum_{\alpha,\beta,\gamma=0}^3
  A_{\mu\alpha}(1,\xv) A_{\nu\beta}(2,\xv') A_{\tau\gamma}(3,\xv'')
  \nonumber\\&\quad\times
  \phizero^{[\alpha+\beta+\gamma]}
  \big(n_3(1,\xv)+n_3(2,\xv')+n_3(3,\xv'')\big).
  \label{EQphiTernary}
\end{align}
Eqs.~(\ref{EQphiTernary}) and (\ref{EQFexcTernaryExplicit}) together
with Eq.~(\ref{EQAterms}) (where $i=1,2,3$) prescribe the DFT for the
general ternary hard sphere mixture. It is straightforward to verify
that this functional reduces, in the corresponding limits, to the FMT
for additive ternary hard sphere mixtures
\cite{rosenfeld89,kierlik90}, and for binary NAHS mixtures
\cite{schmidt04nahs}. The diagrammatical structure of
(\ref{EQFexcTernaryExplicit}) is shown in Fig.\ \ref{FIGdiagrams}d.

\section{Results for bulk fluid structure}
\begin{figure}[htbp]
\centering
\includegraphics[width=6cm,angle=-90]{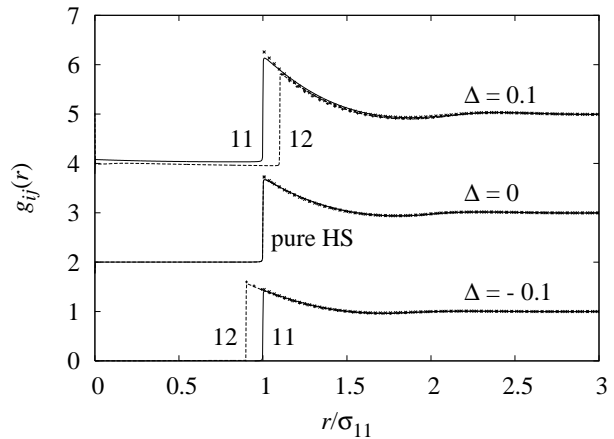}
\caption{Partial pair correlation function $g_{ij}(r)$ for symmetric
  ternary non-additive hard sphere mixtures with
  $\sigma_{11}=\sigma_{22}=\sigma_{33}\equiv \sigma$ for varying
  degree of non-additivity
  $\Delta_{12}=\Delta_{23}=\Delta_{13}\equiv\Delta=-0.1,0,0.1$ as a
  function of the scaled distance $r/\sigma$. Symbols denote Monte
  Carlo data; lines represent DFT results. Shown are the like
  correlation function $g_{11}(r)$ (solid lines and crosses) and the
  unlike correlation function $g_{12}(r)$ (dashed lines and
  pluses). The results for $\Delta=0$ (0.1) are shifted upwards by two
  (four) units.}
\label{FIGgijComparisonOne}
\end{figure}

\begin{figure}[htbp]
\centering
\includegraphics[width=6cm,angle=-90]{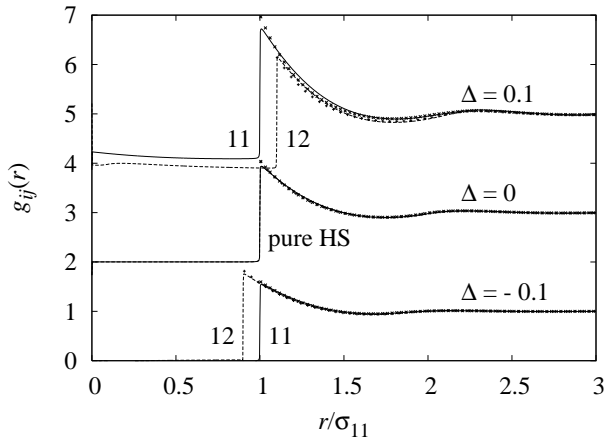}
\caption{Same as Fig.~\ref{FIGgijComparisonOne}, but for total packing
  fraction $\eta=0.25$.}
\label{FIGgijComparisonTwo}
\end{figure}

We test the theory by calculating the partial two-body direct
correlation functions for bulk fluids from $c_{ij}(|\rv-\rv'|) =
-(k_BT)^{-1} \left.  \delta^2 F_{\rm exc}/\delta \rho_i(\rv) \delta
\rho_j(\rv') \right|_{\rho_k=\mathrm{const}}$. Analytic expressions
for the corresponding expressions $\tilde c_{ij}(q)$ in Fourier space
can be obtained. Inserting these into the Ornstein-Zernike equation
for ternary mixtures and Fourier transforming numerically yields
partial pair correlation functions $g_{ij}(r)$.  Such results are
shown in Fig.\ \ref{FIGgijComparisonOne} for symmetric mixtures,
$\sigma_{11}=\sigma_{22}=\sigma_{33}\equiv\sigma$, with varying degree
of non-additivity $\Delta_{12}=\Delta_{23}=\Delta_{23}\equiv \Delta$,
and at equal (and constant) bulk densities, $\rho_1=\rho_2=\rho_3$. We
choose the overall packing fraction as $\eta= \sum_i \pi \rho_i
\sigma_{ii}^3/6=0.2$ and compare to benchmark Monte Carlo simulation
results for 1023 particles, obtained with $10^6$ attempted moves per
particle of which the initial $10^5$ moves were used for
equilibration. Except for (numerically) small core violation, the OZ
results reproduce the simulation data very well. This is also the case
for the slightly higher total packing fraction, $\eta=0.25$, where
again MC and DFT data are compared in Fig.~\ref{FIGgijComparisonTwo}.

In order to consider a fully asymmetric case, we have chosen the
$\sigma_{ij}$ in order to mimick the correlations that were obtained
in Molecular Dynamics simulations of an aqeuous binary salt solution
\cite{kalcher10jcp}. Here the solvent is at large packing fraction and
both types of ions are at vanishing concentration.  In this case the
model parameters can be tuned to mimick the simulation data (not
shown). This constitutes a step towards possible applications of the
current theory to electrolyte solutions. However, for mixtures with
very asymmetric size ratios one would expect to find similar
shortcomings as are present in the additive FMT \cite{herring06prl}
and indeed Percus-Yevick theory.

\section{Conclusions}
In conclusion, we have presented a fundamental measure functional for
ternary NAHS mixtures. The mathematical structure of the functional is
based on a diagrammatic expansion in density with star-like
(tree-like) topology of the diagrams. We have shown that pair
correlation functions obtained from the density functional via the
Ornstein-Zernike route agree well with computer simulation
data. Possible applications of the current theory include its use as a
reference system in modeling electrolytes in bulk and in inhomogeneous
situations. Furthermore, considering one-dimensional cases, see
Ref.\ \cite{santos08oned} for the exact solution of bulk properties of
binary mixtures, could be interesting.

It is worthwhile to discuss possible generalization to mixtures with
more than three components. The general ternary case presented here
rests crucially on the decomposition of the interaction distances
$\sigma_{ij}$ into suitable length scales $R_i$ and $d_i$,
cf.\ (\ref{EQone}). This allowed to use suitable diagrams with
tree-like structure, cf.\ Fig.\ \ref{FIGdiagrams}d. However,
(\ref{EQone}) does not have a general solution (for $R_i$ and $d_i$
once the $\sigma_{ij}$ are prescribed) for $M\geq 4$ components. That
this is true can be gleaned from the fact that the $\sigma_{ij}$ are
$M(M+1)/2$ independent constants, whereas the $R_i$ and $d_i$
constitute only $2M$ free parameters. These numbers, seemingly by
accident, match in the special case $M=3$. Hence we can conclude that
possible extensions to four and more components requires further
changes in the mathematical structure of the present FMT.

\acknowledgments I thank Paul Hopkins for many stimulating discussions
about the physics of non-additive hard spheres. A.J.\ Archer is
acknowledged for useful comments and M. Burgis for independently
observing the upper limits in (\ref{EQ_Phi_munu}). This work was
supported by the EPSRC under Grant EP/E065619/1 and by the DFG via
SFB840/A3.

%% \bibliographystyle{prsty}
%% \bibliography{FMF}

\end{document}